\documentclass[%
 reprint,
 amsmath,amssymb,
 aps,
 prapplied,
]{revtex4-2}

\usepackage{graphicx}
\usepackage{dcolumn}
\usepackage{bm}

\usepackage[utf8]{inputenc}
\usepackage[T1]{fontenc}

\usepackage{mathptmx}
\usepackage{etoolbox}

\makeatletter
\def\@email#1#2{%
 \endgroup
 \patchcmd{\titleblock@produce}
  {\frontmatter@RRAPformat}
  {\frontmatter@RRAPformat{\produce@RRAP{*#1\href{mailto:#2}{#2}}}\frontmatter@RRAPformat}
  {}{}
}%
\makeatother
\begin{document}

\title[Article]{Universal scaling and measurability limits of  thermal-conductivity measurement by a nanoscale point contact}
\author{Ram Munde}
\affiliation{School of Materials Engineering, Purdue University, West Lafayette, Indiana 47907, USA}

\author{Raisul Islam\textsuperscript{*}}
\email{raisul@purdue.edu}
\affiliation{School of Materials Engineering, Purdue University, West Lafayette, Indiana 47907, USA}
\affiliation{Elmore Family School of Electrical and Computer Engineering, Purdue University, West Lafayette, Indiana 47907, USA}

\date{October 17, 2026}

\begin{abstract}
Thermal conductivity ($\kappa$) measurement of a thin film is critical for the development of materials for thermal management in heterogeneous systems. A nanoscale point contact method based on the scanning thermal microscopy (SThM) technique offers a unique approach to probe $\kappa$ with nanoscale spatial resolution in sub-100 nm thick films. Extracting $\kappa$ of a thin film from this method requires separating film, substrate, and interface contributions that are comparable in magnitude. The problem of extracting the right target variable becomes a higher dimensional problem due to multiple unknowns. In this work, we used dimensional analysis to reduce the number of variables in order to provide physical insights into the realistic bounds of $\kappa$ measurement. We show that the total tip-sample thermal resistance $R_{ss}$, originally a function of six variables, reduces to four independent dimensionless groups. These dimensionless groups are used to construct a mathematical relation that captures the underlying dependence of $R_{ss}$ on the governing parameters. The resulting analysis shows that the point contact tip-film-substrate problem is governed by a single dimensionless invariant, $\Omega = R_{ss}\kappa_{\mathrm{sub}}t$, and two control parameters; the confinement ratio $t/b$ and the thermal contrast $\kappa_{\mathrm{sub}}/\kappa_{\mathrm{film}}$. From the invariant, we derive an amplification factor, set only by the contact geometry and the interfaces, that converts the uncertainty of a measurement into an uncertainty in the extracted conductivity. This factor scales with the ratio of the parasitic interface resistance to the film's sheet thermal resistance and bounds the film conductivity that a given contact can resolve. We apply the resulting mathematical framework to a 14$\pm$1 nm Al film grown on SiO$_2$ to locate the measurement within the established bound. Then, the framework allows us to map its thermal resistance with sub-100 nm spatial resolution and recover the local thermal conductivity.

\end{abstract}

\maketitle

\section{\label{sec:1}Introduction}

Thermal conductivity ($\kappa$) is a critical property for material selection in advanced system-technology co-optimization (STCO)~\cite{zhou2026material, pal2024system, zhang2026point, cao2025thermal}. As device dimensions continue to scale down, accurately measuring $\kappa$ of a thin film has become important, yet non-trivial. Established techniques such as
time-domain thermoreflectance (TDTR)~\cite{cahill2004analysis}, frequency-domain
thermoreflectance (FDTR)~\cite{schmidt2009frequency}, and the $3\omega$
method~\cite{cahill1990thermal} have been used successfully for bulk materials and films hundreds of nanometers thick, but they face fundamental limitations in the sub-100~nm regime. In TDTR and FDTR, the thermal resistance of a film thinner
than roughly 80~nm becomes comparable to or smaller than the thermal boundary
resistances at the transducer/film and film/substrate interfaces. Therefore, the measured signal becomes weakly sensitive to the intrinsic $\kappa$ of the film, and
the extracted value is strongly correlated with the interfacial parameters, resulting
 in large uncertainties~\cite{scott2018thermal, song2021thermal}. The $3\omega$ method requires a lithographically patterned metal heater, an
electrically insulating layer, and typically differential measurements against
reference samples of varying thicknesses~\cite{grosse2017development}. For sub-100~nm films, where the film
contributes only a small fraction of the total measured temperature rise, this
sample preparation becomes challenging~\cite{cahill1994thermal, lee1997heat}. These limitations highlight the importance of alternative approaches that probe heat transport locally, with sensitivity that remains high as the film thickness decreases to sub-100 nm regime. 

Nanoscale point-contact techniques offer a route around these limitations. In scanning thermal microscopy (SThM), a heated probe with a tip radius of tens of nanometers is brought into contact with the sample surface, and the
local heat flow from the tip into the sample is inferred from the probe's thermal response measured by reading the change in electrical resistance~\cite{majumdar1999scanning,cahill2014nanoscale} . The probe material is selected based on the temperature dependence of its electrical resistance~\cite{munde2026}. Because the film is locally heated through a nanoscale contact rather than an optical spot, SThM is not bound by the optical diffraction limit, and spatial resolutions approaching 10~nm have been demonstrated~\cite{majumdar1993thermal}.
Accordingly, the heated volume beneath the contact is minuscule, which makes the measurement intrinsically sensitive to the near-surface region. Increasingly, SThM has been used to map local variations in thermal transport associated with microstructural features such as grain boundaries,
dislocations, and buried interfaces~\cite{oesterschulze1996thermal,alikin2024nanoscale}. Extracting a quantitative value of $\kappa$ from these measurements, however,
requires separating the sample's spreading thermal resistance from the tip-sample
contact thermal resistance and parasitic heat-loss channels such as air conduction
and the water meniscus~\cite{zhang2020review,li2025quantitative}. How accurately this separation can be made, and for what range of film thickness,
film conductivity, and interfacial thermal resistance, has not been studied systematically. 

In this work, we develop a scaling framework for point-contact thermal measurements for thin films and use it to determine when such measurements can resolve the film thermal conductivity with nanoscale spatial resolution. Using dimensional analysis based on the Buckingham $\Pi$ theorem~\cite{buckingham1914physically}, we reduce the six variables governing the thermal transport problem to four dimensionless variable groups, yielding a material-independent multi-variable relation. Furthermore, we demonstrate that the tip-film-substrate system is characterized by a single invariant, $\Omega = R_{ss}\kappa_{\mathrm{sub}}t$. Here, $R_{ss}$ is the tip-sample thermal resistance, $\kappa_{\mathrm{sub}}$ is the substrate thermal conductivity, and $t$ is the film thickness. This invariant is governed entirely by the confinement ratio $t/b$ (where $b$ is the contact radius) and the thermal contrast $\kappa_{\mathrm{sub}}/\kappa_{\mathrm{film}}$. From $\Omega$, we
derive a bound on the film conductivity that a given contact can resolve. This factor is set by the ratio of the parasitic interface resistance to the film's sheet thermal resistance. It depends on the contact and the interfaces alone, so it represents an intrinsic limit of the point-contact geometry rather than of any particular instrument. In the high-contrast regime (film thermal conductivity $\gg$ substrate thermal conductivity), the inversion from $R_{ss}$ to $\kappa_{\mathrm{film}}$ becomes explicit, requiring no iterative fitting. We detailed the iterative fitting approach in our previous work~\cite{munde2026}. Finally, we apply the framework to a $14\pm1$~nm Al film on SiO$_2$, locate the measurement within the established bound, and demonstrate thermal conductivity mapping with nanoscale spatial resolution.

The remainder of this paper is organized as follows. Section~\ref{sec:2}
formulates the tip-film-substrate conduction problem and identifies its
governing variables. Section~\ref{sec:3}.A derives the invariant $\Omega$,
constructs the material independent mathematical framework, and presents the resulting universal scaling map.
Section~\ref{sec:3}.B derives the measurability bound and the explicit
inversion in the high-contrast limit. Section~\ref{sec:3}.C applies the
framework to the Al/SiO$_2$ system and presents the thermal conductivity map.
Section~\ref{sec:conclusion} summarizes the results.

\section{\label{sec:2}Theoretical Framework}
\subsection{Point Contact Thermal Resistance Model}
\label{sec:model}

\begin{figure}[t]
    \centering
    \includegraphics[width=0.9\linewidth]{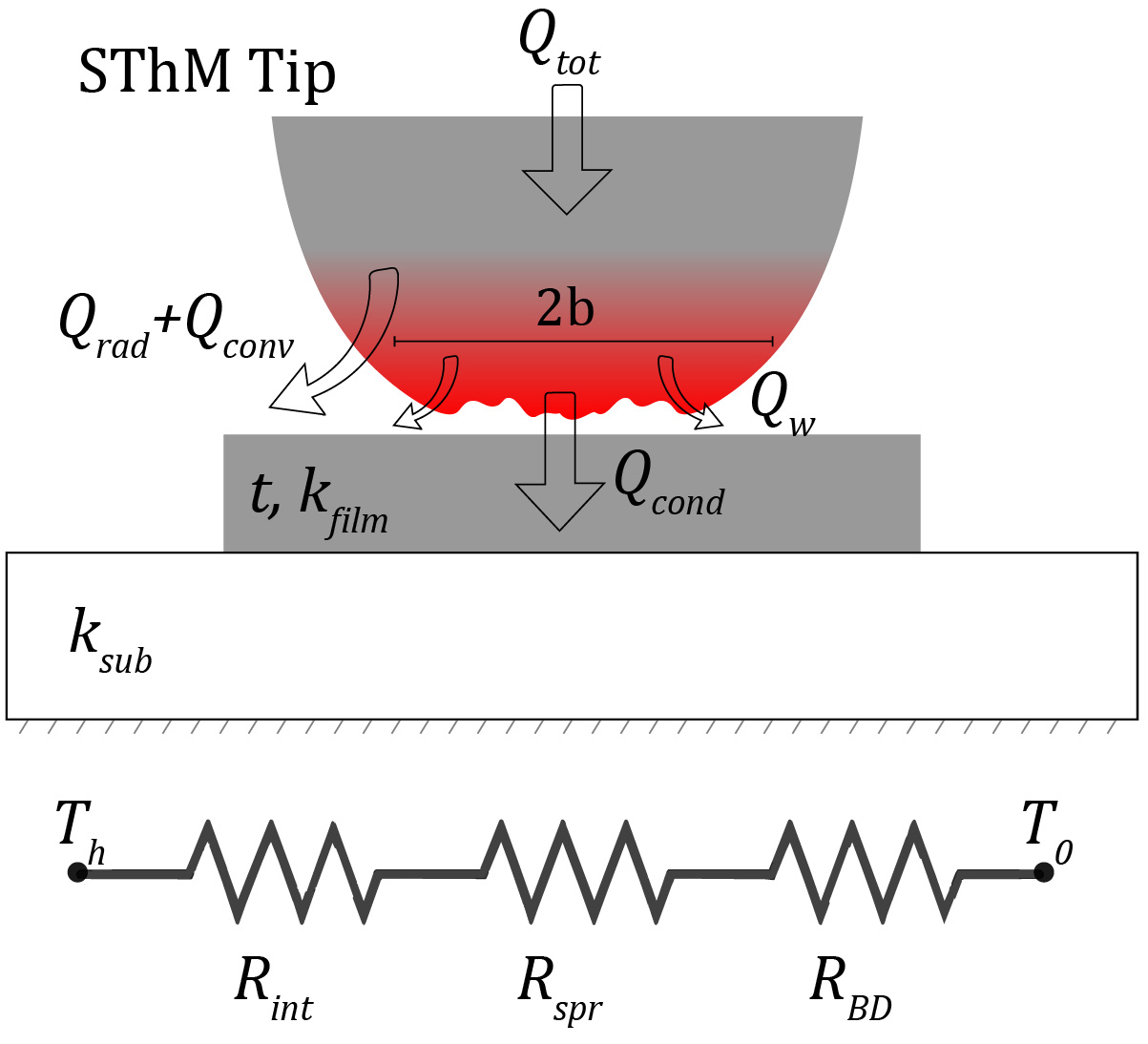}
    \caption{Schematic of the SThM tip-sample contact geometry.}
    \label{fig:network}
\end{figure}

We model the total thermal resistance from tip to sample, $R_{ss}$, as a series network
of three resistances (Fig.~\ref{fig:network}), neglecting radiative and
convective losses~\cite{munde2026, zhang2020review}:
\begin{equation}
\begin{split}
R_{ss} ={}& R_{\mathrm{int}}(h_c, b)
+ R_{\mathrm{spr}}(t, b, \kappa_{\mathrm{film}}, \kappa_{\mathrm{sub}}) \\
&+ f\!\left(\frac{t}{b}, \frac{\kappa_{\mathrm{film}}}{\kappa_{\mathrm{sub}}}\right)
R_{\mathrm{BD}}(h_{\mathrm{BD}}, b)
\end{split}
\label{eq:network}
\end{equation}
Here, $R_{\mathrm{int}}$ is the tip-sample interface resistance,
$R_{\mathrm{spr}}$ is the thermal spreading resistance of the film-substrate stack,
$R_{\mathrm{BD}}$ is the thermal boundary resistance of the buried
film-substrate interface, and $f$ is a geometric dilution factor. Each term is defined below.

\textit{Tip-sample interface resistance.} Heat enters the sample through a contact with a radius of $b$. Therefore, $R_{\mathrm{int}}$ is defined as follows:
\begin{equation}
R_{\mathrm{int}} = \frac{1}{\pi b^2 h_c},
\label{eq:r_int}
\end{equation}
where $h_c$ is the tip-sample interfacial thermal conductance (W/m$^2$$\cdot$K). In general,
$h_c$ depends on surface roughness, tip wear and tear, and the local topography at the
contact, and these effects are absorbed into the effective $h_c$ value~\cite{Kim2008QuantitativeTechnique,zhang2020review}. 

\textit{Thermal spreading resistance.} Beneath the contact, the sample is treated as a
film of thickness $t$ on a semi-infinite substrate (See Fig.~\ref{fig:network}).
For this layered geometry, Dryden~\cite{dryden1983}, Yovanovich, and
Muzychka~\cite{muzychka1999thermal,yovanovich2002analytical} derived the
spreading resistance from  a circular contact with radius b as:
\begin{equation}
R_{\mathrm{spr}} = \frac{1}{\pi \kappa_{\mathrm{film}} b}
\int_{0}^{\infty}
\frac{1 + K e^{-2\xi t/b}}{1 - K e^{-2\xi t/b}}\,
J_1(\xi)\,\sin\xi\,
\frac{d\xi}{\xi^2},
\label{eq:rspr}
\end{equation}
where $J_1$ is the Bessel function of the first kind of order one and
\begin{equation}
K = \frac{\kappa_{\mathrm{film}} - \kappa_{\mathrm{sub}}}
{\kappa_{\mathrm{film}} + \kappa_{\mathrm{sub}}}
\label{eq:K}
\end{equation}
is the conductivity mismatch parameter. Equation~\eqref{eq:rspr} recovers the
half-space result $R_{\mathrm{spr}} = 1/(4\kappa b)$ in both limits:
$\kappa = \kappa_{\mathrm{sub}}$ as $t/b \to 0$, and
$\kappa = \kappa_{\mathrm{film}}$ as $t/b \to \infty$.

\textit{Film-substrate boundary resistance.} The buried interface is
characterized by the thermal boundary conductance $h_{\mathrm{BD}}$, an
intrinsic property of the interface. Its thermal resistance evaluated over the contact area is formulated as:  

\begin{equation}
R_{\mathrm{BD}} = \frac{1}{\pi b^2 h_{\mathrm{BD}}}
\label{eq:r_bd}
\end{equation}
Heat spreads laterally within the film before it reaches the buried
interface, so it crosses that interface over an area larger than $\pi b^2$.
The dilution factor $f$ accounts for this. Using a geometric argument
discussed in Ref.~\cite{munde2026}, we take the effective radius at the
interface to be $b + t\sqrt{\kappa_{\mathrm{film}}/\kappa_{\mathrm{sub}}}$,
which gives
\begin{equation}
f = \left(\frac{b}{b + t\sqrt{\kappa_{\mathrm{film}}/\kappa_{\mathrm{sub}}}}\right)^{2}
\label{eq:f}
\end{equation}
Together, Eqs.~\eqref{eq:network}-\eqref{eq:f} show that $R_{ss}$ depends on
six dimensional variables: $h_c$, $h_{\mathrm{BD}}$, $b$, $t$,
$\kappa_{\mathrm{film}}$, and $\kappa_{\mathrm{sub}}$.

\begin{table*}
\caption{\label{tab:pi_groups}Physical meaning and limitations of the
dimensionless groups governing the point contact thermal resistance of a supported
film.}
\begin{ruledtabular}
\begin{tabular}{c p{0.36\textwidth} p{0.48\textwidth}}
Group & Physical meaning & Limiting behavior \\
\hline
$\Pi_1$
 & Measured thermal resistance in units of the film's own spreading
   scale $1/(\kappa_{\mathrm{film}}b)$.
 & $\Pi_1 = 1/4$ for an isothermal contact on a half space of film
   material, $R_{\mathrm{spr}}=1/(4\kappa_{\mathrm{film}}b)$; any excess over
   $1/4$ arises from effects other than intrinsic film physics. \\[4pt]
$\Pi_2$
 & Geometric confinement ratio; sets how much of the thermal field
   samples the film.
 & $\Pi_2 \gg 1$: heat stays within the film and the substrate is
   screened. $\Pi_2 \ll 1$: heat crosses directly into the substrate
   and the film contributes little to the signal. \\[4pt]
$\Pi_3$
 & Thermal contrast; governs the degree of lateral spreading.
 & $\Pi_3 \ll 1$: the film acts as a heat spreader on a poorly
   conducting support, and heat travels laterally before entering the
   substrate. \\[4pt]
$\Pi_4$
 & Biot number of the buried interface; compares the conductance across
   the interface with conduction through the film.
 & $\Pi_4 \gg 1$: interface is thermally transparent.
   $\Pi_4 \ll 1$: interface is the bottleneck. \\[4pt]
$\Pi_5$
 & Biot number of the tip-sample junction; sets the fraction of the
   measured resistance arising from the contact rather than the sample.
 & Small $\Pi_5$: the measurement is dominated by the tip-sample
   contact. \\[4pt]
$y = \Pi_2/\Pi_3$
 & Spreader strength; combines the film-sensing terms into one
   parameter, the film sheet conductance relative to that of the
   substrate over the contact scale.
 & Large $y$: strong lateral spreading in the film. Small $y$: the
   substrate dominates. \\
\end{tabular}
\end{ruledtabular}
\end{table*}

\subsection{Dimensional analysis: Buckingham $\Pi$ Theorem}
\label{sec:scaling}
Dimensional analysis reduces a physical relation to its minimal set of
independent parameters and has produced familiar groups such as the Reynolds,
Prandtl, and Rayleigh numbers~\cite{white1994buckingham,tamburrino2024old}, as
well as scaling relations in fields ranging from percolation to
biology~\cite{essam1980percolation,mcmahon2003reliability,larkin2018signal,fitzhugh1973dimensional}.
Although it has been used since the nineteenth century, its formalization as
the $\Pi$ theorem is generally attributed to
Buckingham~\cite{buckingham1914physically}.

Equation~\eqref{eq:network} relates the dependent variable $R_{ss}$ to six independent variables:
$\{\kappa_{\mathrm{film}}, \kappa_{\mathrm{sub}}, b, t, h_{\mathrm{BD}}, h_c\}$. These involve length ($L$), temperature ($\Theta$), and power ($W$), but $\Theta$ and $W$ appear only in the combination $\Theta W^{-1}$: thermal resistance has dimensions $\Theta W^{-1}$, thermal conductivity
$L^{-1}(\Theta W^{-1})^{-1}$, and thermal conductance
$L^{-2}(\Theta W^{-1})^{-1}$. Therefore, $L$ and $\Theta W^{-1}$ are the only linearly independent variables. Thus, the dimensional matrix has rank $D = 2$, and the $\Pi$ theorem gives $6 - D = 4$ independent dimensionless groups. Taking $\kappa_{\mathrm{film}}$ and $b$ as repeating variables~\cite{buckingham1914physically}, we obtain
\begin{equation}
\Pi_2 = \frac{t}{b}, \quad
\Pi_3 = \frac{\kappa_{\mathrm{sub}}}{\kappa_{\mathrm{film}}}, \quad
\Pi_4 = \frac{h_{\mathrm{BD}} b}{\kappa_{\mathrm{film}}}, \quad
\Pi_5 = \frac{h_c b}{\kappa_{\mathrm{film}}},
\label{eq:pis}
\end{equation}
with the dependent group $\Pi_1 = R_{ss}\kappa_{\mathrm{film}} b$. Here,
$\Pi_2$ is the confinement ratio, $\Pi_3$ is the thermal conductivity contrast between the film and the substrate, and
$\Pi_4$ and $\Pi_5$ are the Biot numbers of the buried interface and the tip
contact, respectively. The complete derivation of $\Pi$ groups is given in the Supplementary
Information~\cite{SM}.

Multiplying Eq.~\eqref{eq:network} by $\kappa_{\mathrm{film}} b$ and using
Eqs.~\eqref{eq:r_int} and \eqref{eq:r_bd}, the interface terms become
\begin{equation}
R_{\mathrm{int}}\kappa_{\mathrm{film}} b = \frac{1}{\pi\Pi_5}, \qquad
R_{\mathrm{BD}}\kappa_{\mathrm{film}} b = \frac{1}{\pi\Pi_4},
\label{eq:Rbd_pi}
\end{equation}
and the Eq.~\eqref{eq:network} takes the dimensionless form as:
\begin{equation}
\boxed{\;
\Pi_1 = \frac{1}{\pi\Pi_5} + \Psi(\Pi_2, \Pi_3)
+ \frac{f(\Pi_2, \Pi_3)}{\pi\Pi_4}
\;}
\label{eq:master_explicit}
\end{equation}
The dilution factor [Eq.~\eqref{eq:f}] becomes
$f = \left(1 + \Pi_2\Pi_3^{-1/2}\right)^{-2}$, and the dimensionless spreading
resistance is
\begin{equation}
\Psi(\Pi_2, \Pi_3) \equiv R_{\mathrm{spr}}\kappa_{\mathrm{film}} b
= \frac{1}{\pi}\int_{0}^{\infty}
\frac{1 + K e^{-2\xi\Pi_2}}{1 - K e^{-2\xi\Pi_2}}\,
\frac{J_1(\xi)\sin\xi}{\xi^{2}}\,d\xi,
\label{eq:psi}
\end{equation}
where $K = (1 - \Pi_3)/(1 + \Pi_3)$ is from Eq.~\eqref{eq:K}. Because both $K$
and the exponent depend only on $\Pi_2$ and $\Pi_3$, $\Psi$ is a universal
function of these two groups: any combination of film, substrate, and probe
sharing the same $(\Pi_2, \Pi_3)$ has the same dimensionless spreading
resistance. Within the assumptions of the network model of Eq.~\eqref{eq:network},
Eq.~\eqref{eq:master_explicit} is therefore a complete dimensionless
description of the tip-sample thermal resistance.

\begin{figure*}[t]
\includegraphics[width=\textwidth]{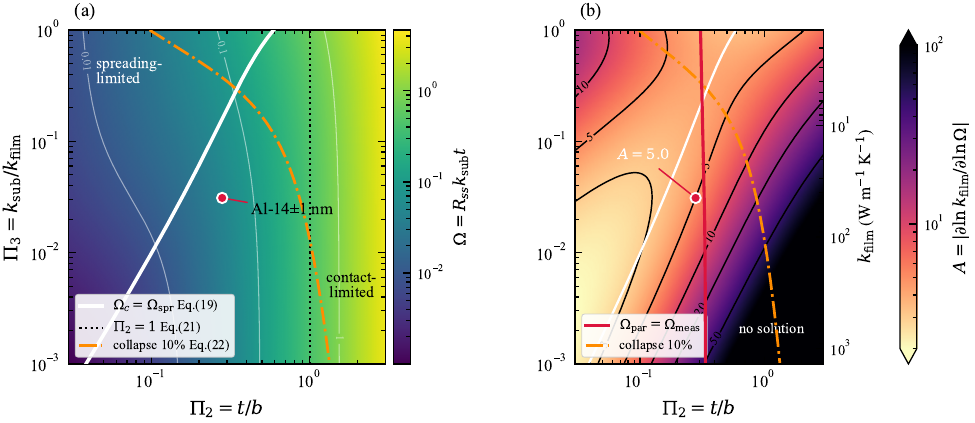}
\caption{Regime map at the nominal film thickness $t = 15$\,nm (calibrated parameters for Al film: $\Lambda_c = 0.753$,
$\Lambda_{\mathrm{BD}} = 2.17$); varying $\Pi_2$ corresponds to varying $b$.
(a)~Invariant $\Omega$ (color). Solid white: contact/spreading boundary
[Eq.~\eqref{eq:bd_contact}]; dotted: $\Pi_2 = 1$ [Eq.~\eqref{eq:bd_thick}];
dash-dotted: $10\%$ collapse boundary [Eq.~\eqref{eq:bd_collapse}]. The
buried-interface boundary [Eq.~\eqref{eq:bd_bdi}] has no solution here.
(b)~Amplification factor $A$ [Eq.~\eqref{eq:amplification}] (color); right
axis gives $\kappa_{\mathrm{film}}$. white (solid line): Eq.~(\ref{eq:bd_contact}) from panel (a). red: the locus
$\Omega_{\mathrm{par}} = \Omega_{\mathrm{meas}}$ [Eq.~\eqref{eq:solvability}], which bounds the region containing the solution and asymptotes to $\Pi_2$ = 0.357 as $\Pi_3\rightarrow0$; higher than that contact resistance exceeds the measurement and no solution exists for any film.
Circles mark the Al film at $(\Pi_2,\Pi_3) = (0.28,\,0.031)$ ($\kappa_{\mathrm{film}} \approx 44.5$\,W\,m$^{-1}$K$^{-1}$, from inverting the measured invariant $\Omega_{\mathrm{meas}}$; the 37.9\,W\,m$^{-1}$K$^{-1}$ of Fig.~\ref{fig:kmap}(c) is the median of the log-normal fit to the pixel distribution), placed from its measured thickness $t=14\pm1$\,nm~\cite{munde2026} and b= 50 nm; the difference from the nominal thickness lies within the measurement uncertainty. The plotting code for this figure was written with the assistance of Claude Opus 5 (Anthropic) from the authors' expressions [Eqs.~\eqref{eq:omega}, \eqref{eq:bd_contact}, \eqref{eq:bd_thick}, \eqref{eq:bd_collapse}, \eqref{eq:solvability} and \eqref{eq:amplification}]. The plotted curves were verified against those expressions.}
\label{fig:surface}
\end{figure*}

\section{\label{sec:3}Results and Discussion}

The six-variable SThM tip-sample thermal resistance problem is now transformed into four independent dimensionless groups $\Pi_i$, along with the dependent group $\Pi_1$. Each of these carries a direct physical meaning, as described in Table~\ref{tab:pi_groups}.

\subsection{Regime structure}

The essential simplification for thermal conductivity measurement follows by fixing the film thickness $t$ and depositing a uniform-thickness film on a substrate with known $\kappa_\mathrm{sub}$. Then $b = t/\Pi_2$, so $\Pi_4$ and $\Pi_5$ are no longer free:
\begin{align}
&\Pi_4=\Lambda_{\mathrm{BD}}\frac{\Pi_3}{\Pi_2},\qquad
\Pi_5=\Lambda_{c}\frac{\Pi_3}{\Pi_2},\nonumber\\
&\Lambda_{\mathrm{BD}}=\frac{h_{\mathrm{BD}}t}{k_{\mathrm{sub}}},\qquad
\Lambda_{c}=\frac{h_ct}{k_{\mathrm{sub}}},
\label{eq:slaved}
\end{align}
where $\Lambda_{\mathrm{BD}}$ and $\Lambda_c$ are Biot numbers corresponding to the film-substrate and tip-sample interfaces, respectively~\cite{biot}. These are material specific constants that can be determined through independent experiments~\cite{munde2026,reifenberg2009thermal}. Next, multiplying
Eq.~(\ref{eq:master_explicit}) by $\Pi_2\Pi_3$ eliminates both unknowns from the left-hand side because
$\Pi_1\Pi_2\Pi_3=R_{ss}\kappa_{\mathrm{sub}}t$ contains neither $b$ nor $\kappa_{\mathrm{film}}$. We treat $b$ as an independent variable rather than the probe apex radius since,
 at the nanoscale, the thermal contact radius may differ from the
physical tip radius. Consequently, defining the
measured invariant $\Omega\equiv R_{ss}\kappa_{\mathrm{sub}}t$, it gives:
\begin{equation}
\Omega=\mathcal{H}(\Pi_2,\Pi_3)
=\frac{\Pi_2^{2}}{\pi}\left(\frac{1}{\Lambda_{c}}+\frac{f}{\Lambda_{\mathrm{BD}}}\right)
+\Pi_2\Pi_3\,\Psi(\Pi_2,\Pi_3).
\label{eq:omega}
\end{equation}
This is the central equation of this problem. At fixed $t$ and Biot numbers $\Lambda_{\mathrm{BD}}$ and $\Lambda_c$ , the thermal conductivity depends on a single two-parameter mathematical framework, as shown in Fig.~\ref{fig:surface}(a). In Fig.~\ref{fig:surface}(a), $\Pi_2$ encodes the local contact radius, $\Pi_3$ the local thermal conductivity, and the SThM based point contact measurement fixes the value of $\Omega$. Therefore, once $\Pi_2$ is known, $\mathcal{H}$ is monotonically increasing in $\Pi_3$, ensuring a unique solution for the local thermal conductivity (see Fig.~S1 in the Supplementary Information). Importantly, apart from the weak $\Pi_3$ dependence of $f$, this formulation eliminates the need for the iterative solution~\cite{munde2026}, enabling the extraction of thermal conductivity in a single vectorized operation.

\subsubsection{Decomposition of the invariant}

Equation~\eqref{eq:omega}  can be expressed as the invariant with a sum of three positive
contributions,
\begin{equation}
\Omega = \underbrace{\frac{\Pi_2^{2}}{\pi\Lambda_c}}_{\Omega_c}
\;+\; \underbrace{\frac{f\Pi_2^{2}}{\pi\Lambda_{\mathrm{BD}}}}_{\Omega_{\mathrm{BD}}}
\;+\; \underbrace{\Pi_2\Pi_3\,\Psi(\Pi_2,\Pi_3)}_{\Omega_{\mathrm{spr}}},
\label{eq:omega_split}
\end{equation}
representing the variables associated with tip-sample contact ($\Omega_c$), the buried film-substrate interface ($\Omega_\mathrm{BD}$),
and the heat spreading within the film-substrate stack ($\Omega_\mathrm{spr}$). To a good approximation, only $\Omega_{\mathrm{spr}}$
carries information about $\kappa_{\mathrm{film}}$; $\Omega_{\mathrm{BD}}$ retains a weak dependence through $f$. We therefore treat the first two terms as
parasitic and group them as
$\Omega_{\mathrm{par}} \equiv \Omega_c + \Omega_{\mathrm{BD}}$. The
$(\Pi_2,\Pi_3)$ axes in Fig.~\ref{fig:surface}(a) therefore divide the $\Omega$ mapping plane into regions according to which term dominates, and the boundaries are the loci on which pairs of terms are equal.
\subsubsection{Exact limits of the spreading function}
The boundaries are set by the asymptotic behavior of $\Psi$, which we will now derive in closed form under certain assumptions. For a contact radius much smaller than the film thickness, the
substrate is not sensed, $K$ is irrelevant, and the spreading term takes the form of the isothermal-disk on a half-space of film material.
\begin{equation}
\Psi \to \tfrac{1}{4}, \qquad
R_{\mathrm{spr}} \to \frac{1}{4\kappa_{\mathrm{film}}b}
\qquad (\Pi_2 \to \infty),
\label{eq:psi_hi}
\end{equation}
In the opposite limit ($t$ $\ll$ $b$), the film is thermally transparent, and the same result holds for the substrate. 
\begin{equation}
\Psi \to \frac{1}{4\Pi_3}, \qquad
R_{\mathrm{spr}} \to \frac{1}{4\kappa_{\mathrm{sub}}b}
\qquad (\Pi_2 \to 0),
\label{eq:psi_lo}
\end{equation}
Both follow from
$\int_0^{\infty} J_1(\xi)\sin\xi\,\xi^{-2}d\xi = \pi/4$. Between these limits, for $\Pi_3 \ll 1$ with $\Pi_2 \lesssim 1$, the
kernel of Eq.~\eqref{eq:psi} may be expanded as
$1 - Ke^{-2\xi\Pi_2} \simeq 2\Pi_3(1 + y\xi)$ (section S5 of Supplementary Information), so that $\Psi$ collapses onto a
one-parameter family,
\begin{equation}
4\Pi_3\Psi \simeq \mathcal{G}(y)
= \frac{4}{\pi}\int_{0}^{\infty}
\frac{J_1(\xi)\sin\xi}{\xi^{2}\,(1+y\xi)}\,d\xi,
\qquad
y \equiv \frac{\Pi_2}{\Pi_3} = \frac{\kappa_{\mathrm{film}}t}{\kappa_{\mathrm{sub}}b},
\label{eq:Gdef}
\end{equation}
with $\mathcal{G}(0)=1$. The collapse parameter has a direct physical interpretation:
$y = L/b$, where $L = \kappa_{\mathrm{film}}t/\kappa_{\mathrm{sub}}$ is the
lateral distance over which heat travels in the film before leaking into the
substrate. Because $J_1(\xi)\sin\xi/\xi^2 \to 1/2$ as $\xi\to0$, replacing
$1/(1+y\xi)$ with $1/(y\xi)$ diverges logarithmically, and the correct large-$y$
behavior is given by:
\begin{equation}
\mathcal{G}(y) \simeq \frac{2}{\pi}\,\frac{\ln y + C}{y},
\qquad C = 0.283,
\qquad y \gg 1,
\label{eq:Gasym}
\end{equation}
Equation~\eqref{eq:Gasym}
agrees with Eq.~\eqref{eq:Gdef} within $10\%$ for $y \ge 4.2$, with the
error decreasing monotonically as $y$ increases (see S5. of Supplementary Information). In dimensional form, Eq.~\eqref{eq:Gasym} can be written as,
\begin{equation}
R_{\mathrm{spr}} \simeq
\frac{1}{2\pi\kappa_{\mathrm{film}}t}
\left[\ln\!\left(\frac{\kappa_{\mathrm{film}}t}{\kappa_{\mathrm{sub}}b}\right) + C\right],
\qquad \kappa_{\mathrm{film}}t \gg \kappa_{\mathrm{sub}}b,
\label{eq:sheet}
\end{equation}
It represents the spreading resistance of a conducting sheet having a sheet conductance
$\kappa_{\mathrm{film}}t$, where heat is leaking into an insulating half-space over a lateral
length $\sim L$. Equation~\eqref{eq:sheet} is the thermal analog of the
logarithmic spreading resistance of a thin conducting film with a point contact heat source, and $y \sim 1$ marks the crossover from substrate-dominated to
sheet-dominated spreading. We note that $\mathcal{G}$ is not a power law; any effective exponent fitted to Eq.~\eqref{eq:Gdef} over a finite window is an interpolation of the logarithm in Eq.~\eqref{eq:Gasym} and carries no
scaling significance.

\subsubsection{Regime boundaries}

Equating pairs of terms in Eq.~\eqref{eq:omega_split} provides,
\begin{align}
\text{contact / spreading:} &&
\frac{\Pi_2}{\pi\Lambda_c} &= \Pi_3\,\Psi(\Pi_2,\Pi_3),
\label{eq:bd_contact}\\[2pt]
\text{buried interface / spreading:} &&
\frac{f\,\Pi_2}{\pi\Lambda_{\mathrm{BD}}} &= \Pi_3\,\Psi(\Pi_2,\Pi_3),
\label{eq:bd_bdi}\\[2pt]
\text{film half-space / thin sheet:} &&
\Pi_2 &\sim 1 \quad (t \sim b),
\label{eq:bd_thick}\\[2pt]
\text{collapse validity:} &&
\Pi_3 &\ll 1 \ \text{and}\ \Pi_2 \lesssim 1,
\label{eq:bd_collapse}
\end{align}

Equation~\eqref{eq:bd_contact} is the boundary of principal practical
importance since it fixes the range of conductivities a given point contact can
resolve. High tip-sample interfacial thermal resistance and low film spreading resistance both increase measurement uncertainty. Equation~\eqref{eq:bd_bdi} has no solution for a metal film on an oxidized silicon substrate (Si/SiO$_2$), such
as the Al film studied in Sec.~\ref{sec:app}. However, a buried-interface-limited regime is expected when $\Lambda_{\mathrm{BD}} = h_{\mathrm{BD}}t/\kappa_{\mathrm{sub}}$
is small. It is similarly true for atomically thin films with weak van der Waals interfaces on
high-conductivity substrates, such as MoS$_2$ on Si~\cite{lu2025characterization}. Using Eqs.~\eqref{eq:Gdef} and~\eqref{eq:Gasym}, both
$\Omega_{\mathrm{BD}}$ and $\Omega_{\mathrm{spr}}$ scale linearly in $\Pi_3$ at
small $\Pi_3$. Therefore, the ratio can be written as:
\begin{equation}
\frac{\Omega_{\mathrm{spr}}}{\Omega_{\mathrm{BD}}}
\to \frac{\Lambda_{\mathrm{BD}}}{2}\left(\ln y + C\right),
\label{eq:bdi_ratio}
\end{equation}
This ratio exceeds unity for $y \gtrsim 2$ when
$\Lambda_{\mathrm{BD}} = 2.17$. This $\Lambda_{\mathrm{BD}}$ value is validated in Ref.~\cite{munde2026} for the Al/SiO$_2$ system.  Equation~(\ref{eq:bdi_ratio}) is asymptotic in $y$; however, the threshold it implies, $\Lambda_{\mathrm{BD}}={2}/{(\ln y + C)}$, falls from 2.1 at $y$ = 2 to 0.4 at y = 100; only the latter lies within its range of validity. Searching $10^{-3} \le \Pi_2 \le 10$ and $10^{-6} \le \Pi_3 \le 10^{2}$, we find
$\Omega_{\mathrm{BD}}/\Omega_{\mathrm{spr}} \le 0.17$ throughout and
$\Omega_{\mathrm{BD}}/\Omega \le 0.07$. The buried interface is therefore never
the dominant resistance for a metal film on an SiO$_2$ substrate. Therefore, $h_{\mathrm{BD}}$
does not limit the measurement in this case. Since $\Omega_{\mathrm{BD}}$ scales as $1/\Lambda_{\mathrm{BD}}$, while $\Omega_{\mathrm{spr}}$ does not, that bound places the buried-interface-limited regime at 
$\Lambda_{\mathrm{BD}} = h_{\mathrm{BD}}t/\kappa_{\mathrm{sub}} \lesssim 0.4$; this is true for significantly thinner films with poorer interfaces (e.g., a rough interface).

Equation~\eqref{eq:bd_thick} separates the two limits of
Eqs.~\eqref{eq:psi_hi} and \eqref{eq:psi_lo}: since heat injected through a contact of radius $b$ is sensed to a depth of order $b$, $\Pi_2$ is the ratio of the film thickness to the sensed depth. The crossover is broad, and its width grows with the thermal contrast between the film and the substrate. Finally, Eq.~\eqref{eq:bd_collapse} bounds the validity of the high-contrast collapse rather than separating two dominant terms.

Figure~\ref{fig:surface}(a) shows the resulting universal scaling map. The color scale is
$\Omega$ from Eq.~\eqref{eq:omega}, evaluated at fixed film thickness, so that
motion along the $\Pi_2$ axis corresponds to varying the contact radius. The $\Pi_3$ axis represents the variable $\kappa_\mathrm{film}$ with a constant $\kappa_\mathrm{sub}$ = 1.38 W m$^{-1}$K$^{-1}$. The solid white line represents Eq.~\eqref{eq:bd_contact}, which separates spreading and contact resistance limited regimes. To the
upper left of the white solid line, spreading dominates, and the measurement is sensitive to the
film's thermal conductivity; to its lower right, the contact resistance dominates and
$\Omega \to \Omega_c \propto \Pi_2^{2}$, independent of the film. The dotted line marks Eq.~\eqref{eq:bd_thick}, and the dash-dotted line represents the $10\%$ collapse given in Eq.~\eqref{eq:Gasym} with the boundary conditions mentioned in Eq.~\eqref{eq:bd_collapse}. The two regimes are arranged counter-intuitively with respect to contact size: sharpening the probe (increasing $\Pi_2$) moves the measurement \emph{into} the contact-limited region because $\Omega_c$ grows as $\Pi_2^{2}$, while the film term grows only logarithmically, so the film's contribution to the measurement signal falls. On the other hand, for large contacts (\(\Pi_2 \ll 1\)), the substrate dominates the spreading resistance, making the film nearly thermally invisible. In both cases, the film's contribution becomes too small to measure reliably. Therefore, accurate measurements are possible only within an intermediate range of contact radii.
\subsection{Measurability}
\label{sec:measurability}
\subsubsection{Solvability}
A measured invariant admits a solution for
$\kappa_{\mathrm{film}}$ only if it exceeds the parasitic floor (i.e. $\Omega_{\mathrm{spr}} > 0$), it follows as:
\begin{equation}
\Omega > \Omega_{\mathrm{par}}
= \frac{\Pi_2^{2}}{\pi}
\left(\frac{1}{\Lambda_c} + \frac{f}{\Lambda_{\mathrm{BD}}}\right)
\iff
R_{ss} > R_{\mathrm{int}} + f R_{\mathrm{BD}}
\label{eq:solvability}
\end{equation}
Since $f$ depends on the unknown $\Pi_3$, Eq.~(\ref{eq:solvability}) holds at the solution but cannot be evaluated in advance of it. The exact criterion follows from the limit $\kappa_{\mathrm{film}} \rightarrow \infty$: a perfectly conducting film spreads the injected heat over an unbounded area, so the spreading resistance vanishes and the flux at the buried interface is diluted away, $\Pi_3\Psi \rightarrow 0$ and $f \rightarrow 0$. What remains is the tip-sample contact, which the heat must cross before it can spread at all. Since $\mathcal{H}$ increases monotonically in $\Pi_3$, $\Omega_\mathrm{c}$ is its infimum, and a solution exists if and only if 
\begin{equation}
    \Omega > \Omega_\mathrm{c} = \frac{\Pi_2^2}{\pi\Lambda_\mathrm{c}} \iff R_{ss} > R_\mathrm{int}
    \label{eq:solvability_2}
\end{equation}
The distinction between Eq.~\eqref{eq:solvability} and Eq.~\eqref{eq:solvability_2} is physical. The tip-sample contact thermal resistance is unavoidable: it lies in series ahead of the spreading network. The buried interface can be reduced because the film itself sets the area over which it is crossed, so a better spreader dilutes $R_\mathrm{BD}$ toward zero. Therefore, only the contact term sets the floor.



The uniqueness of the inversion of Eq.~\eqref{eq:sheet}, which follows from the monotonicity of
$\mathcal{H}$ in $\Pi_3$, does not imply that it is well conditioned for $\kappa_{\mathrm{film}}$ extraction. Since
$\kappa_{\mathrm{film}} = \kappa_{\mathrm{sub}}/\Pi_3$, the relevant quantity
is the amplification factor ($A$), which is defined as,
\begin{equation}
A \equiv
\left|\frac{\partial \ln \kappa_{\mathrm{film}}}{\partial \ln \Omega}\right|
= \left|\frac{\partial \ln \Omega}{\partial \ln \Pi_3}\right|^{-1}
\simeq \frac{\Omega}{\Pi_2\left|\partial(\Pi_3\Psi)/\partial\ln\Pi_3\right|},
\label{eq:amplification}
\end{equation}
In other words, A is the factor by which a fractional error in the measured resistance is magnified
in the recovered conductivity. The first two expressions are exact in Eq.~(\ref{eq:amplification}); the third neglects the weak dependence of $\Pi_3$ in $f$. Within the collapse regime of Eq.~\eqref{eq:bd_collapse}, where
$\Pi_3\Psi = \mathcal{G}(y)/4$, the derivative in
Eq.~\eqref{eq:amplification} follows from $\partial (\Pi_3\Psi)/\partial\ln\Pi_3 = -y\mathcal{G}'(y)/4$
at fixed $\Pi_2$, and $A$ takes the following closed form,
\begin{equation}
A = \frac{4\,\Omega_{\mathrm{par}}/\Pi_2 + \mathcal{G}(y)}
{\left|y\,\mathcal{G}'(y)\right|}.
\label{eq:A_collapse}
\end{equation}

\begin{figure*}[htbp]
\includegraphics[width=0.9\textwidth]{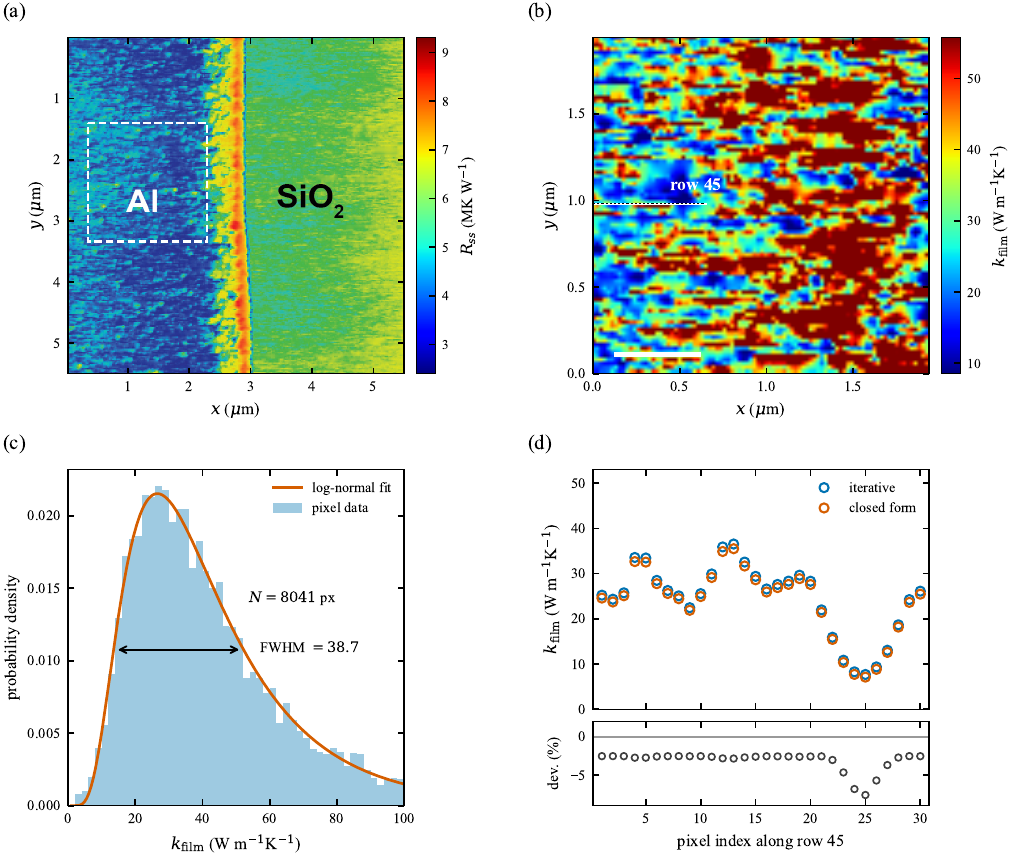}
\caption{\label{fig:kmap} Pixel-level thermal conductivity mapping of a 14$\pm$1 nm Al film on
SiO$_2$(1500 nm)/Si. (a) Measured solid-solid thermal resistance $R_{ss}$ across the
Al/SiO$_2$ edge, from Ref.~\onlinecite{munde2026}; the dashed square marks a region of the
Al film of the size analyzed in (b)-(d). (b) Thermal conductivity map of the Al film,
$1.94\times1.94$ $\mu$m$^2$ (90$\times$90 pixels), obtained by applying the closed-form
inversion, Eq.~(\ref{eq:closed}), to each pixel of the measured $R_{ss}$ map with
$R_{\rm int}=1.84$ MK W$^{-1}$, $b=50$ nm, $t=14$ nm, $\kappa_{\rm sub}=1.38$ W m$^{-1}$K$^{-1}$
and $h_{\rm BD}=200$ MW m$^{-2}$K$^{-1}$. The dashed line indicates the profile analyzed in
(d). (c) Distribution of the extracted conductivities over the $N=8041$ pixels for which
$0<\mathcal{G}<1$, with a log-normal fit giving a median of
\textbf{37.9 W m$^{-1}$K$^{-1}$} and a full width at half maximum of 38.7 W m$^{-1}$K$^{-1}$; the
skew arises because $\kappa_{\rm film}$ varies as $1/\mathcal{G}$, so symmetric noise in $R_{ss}$
maps onto an asymmetric spread in conductivity. (d) Pixel-by-pixel comparison of the
closed-form inversion (Eq.\ref{eq:closed}) with the full iterative solution of Ref.~\cite{munde2026} along
the marked profile, together with the relative deviation (lower panel). The plotting and analysis code for panels (b)-(d), including the log-normal fit in (c) and the iterative comparison in (d), was written with the assistance of Claude Opus 5 (Anthropic). The outputs were verified against the measured data and Eq.~\eqref{eq:closed}.}
\end{figure*}

\subsubsection{The measurability bound}
\label{sec:bound}
Using Eq.~\eqref{eq:Gasym}, the denominator in Eq.~\eqref{eq:A_collapse} can be written as $|y\mathcal{G}'(y)| \simeq (2/\pi)(\ln y + C - 1)/y$
for $y \gg 1$. Assuming the parasitic term dominates the numerator of Eq.~\eqref{eq:A_collapse}, which gives
\begin{equation}
A \simeq \frac{2\pi\,\Omega_{\mathrm{par}}\,y}{\Pi_2\,(\ln y + C - 1)}
= \frac{2\pi\,R_{\mathrm{par}}\,\kappa_{\mathrm{film}}t}{\ln y + C - 1},
\qquad
R_{\mathrm{par}} = R_{\mathrm{int}} + f R_{\mathrm{BD}}.
\label{eq:A_asym}
\end{equation}
Equation~\eqref{eq:A_asym} is the central result of this section. Apart from the slowly varying factor $2\pi/(\ln y + C - 1)$, the amplification is the ratio of the parasitic series thermal resistance to
the sheet thermal resistance $1/(\kappa_{\mathrm{film}}t)$ of the film. The
measurement loses sensitivity as soon as $R_{\mathrm{par}}$ exceeds
$1/(\kappa_{\mathrm{film}}t)$, and it does so linearly in
$\kappa_{\mathrm{film}}$. We find Eq.~\eqref{eq:A_asym} reproduces the exact
$A$ of Eq.~\eqref{eq:amplification} to within $3\%$ for $y = 9$-$30$ and
within $12\%$ up to $y = 300$ for our calibrated data from Ref.~\cite{munde2026}. It does not apply below $y \approx 3$, where the logarithm is not yet dominant. Therefore, for $y \leq 3$ a full model solution must be used. 

Inverting for a chosen tolerance $A_{\max}$ gives the maximum film conductivity
that a given contact can resolve.
\begin{equation}
\kappa_{\mathrm{film}}^{\max}
\simeq \frac{A_{\max}\left(\ln y_{\max} + C - 1\right)}
{2\pi\,R_{\mathrm{par}}\,t},
\qquad
y_{\max} = \frac{\kappa_{\mathrm{film}}^{\max}t}{\kappa_{\mathrm{sub}}b},
\label{eq:kmax}
\end{equation}
The amplification factor $A$ is a property of the contact geometry and the interfaces alone. It contains no reference to detector noise, averaging time, or thermometry. The instrument enters only through the choice of $A_{\max}$, which is the tolerated error in $\kappa_{\mathrm{film}}$ divided by the fractional uncertainty in $R_{ss}$. Better instrumentation, therefore, raises $\kappa_{\mathrm{film}}^{\max}$. Also, note that the bound is one-sided. As $\kappa_{\mathrm{film}}$ decreases, the film's own
resistance dominates $\Omega$ and $A \to 1$; at $b = 50$\,nm, $A$ falls to a plateau of $\simeq 4.1$ over $\kappa_{\mathrm{film}}$ = 2-10 W\,m$^{-1}$K$^{-1}$ and drops below 4 only for $\kappa_{\mathrm{film}} \lesssim 1.9$ W\,m$^{-1}$K$^{-1}$. Therefore, low-conductivity films are comparatively well conditioned, but they lie outside the collapse regime: the $10\%$ collapse boundary at $\Pi_2 = 0.3$ falls at
$\kappa_{\mathrm{film}} = 4.0$\,W\,m$^{-1}$K$^{-1}$. Therefore, below this limit, $\kappa_{\mathrm{film}}$ must be obtained from the full model rather than the explicit inversion.

For a 15\,nm film probed through a point contact of 50\,nm apex radius with $R_{\mathrm{par}} = 1.93$\,MK\,W$^{-1}$ (calibrated in Ref.~\cite{munde2026}), films with the bulk thermal conductivities (\emph{i.e} ., upper bound) of
Al, Au, and Cu (237, 318, and 401\,W\,m$^{-1}$K$^{-1}$) would be recovered
with $A = 12$, $15$, and $17$, so a $1\%$ uncertainty in $R_{ss}$ becomes a
$12$-$17\%$ uncertainty in $\kappa_{\mathrm{film}}$. Point-contact $\kappa$ extraction of
sub-100\,nm metal films is thus limited not by thermometry but by the ratio of
the contact resistance to the sheet thermal resistance of the film, and
extending it to high-conductivity metals requires reducing $R_{\mathrm{par}}$,
primarily by increasing the contact conductance $h_c$.

\subsubsection{Sensitivity and design rule}
\label{sec:sensitivity}


Equation~\eqref{eq:A_asym} yields a design rule. In SThM measurements, the tip-sample contact dominates the parasitic resistance, $R_{\mathrm{par}} \simeq R_{\mathrm{int}} = 1/(\pi b^2 h_c)$. Thus, Eq.~(\ref{eq:A_asym}) becomes,

\begin{equation}
    A \simeq \frac{2\pi}{\ln y+C-1}\frac{\kappa_{\mathrm{film}}t}{\pi b^2 h_c}
    \label{eq:A_int}
\end{equation}
Apart from the slowly varying pre-factor $2\pi/(\ln y+C-1)$, this quantity represents the ratio of the film sheet conductance to the total contact conductance $\pi b^2 h_c$.  Therefore, the film's thermal conductivity is resolvable only when the contact conductance is large compared with $\kappa_{\mathrm{film}}t$. Within the validity range of Eq.~(\ref{eq:A_asym}), $A$ decreases with increasing $h_c$, decreases with increasing $b$, and increases with increasing $t$. The contact radius sets a trade-off between solvability and spatial resolution. Therefore, increasing $h_c$ is the only route that improves the solution space without a resolution penalty.

Figure~\ref{fig:surface}(b) shows $A$ over the same plane ($\Pi_2$, $\Pi_3$), with the right-hand axis giving $\kappa_{\mathrm{film}}$ directly. Here, $\kappa_{\mathrm{film}}$ depends on the control parameters $\Pi_2$ and $\Pi_3$.  Contours of constant $A$ run diagonally: at a fixed contact radius, $A$ grows with $\kappa_{\mathrm{film}}$
, as described in Eq.~\eqref{eq:A_asym}; and at a fixed $\kappa_{\mathrm{film}}$, it
grows with $\Pi_2$ as the contact term takes over. The red curve is the locus $\Omega_{\mathrm{par}} = \Omega_{\mathrm{meas}}$ of Eq.~(\ref{eq:solvability}) evaluated for the measured invariant discussed in the next section (Sec.~\ref{sec:app}). To the right of this locus line, the parasitic resistance dominates, which makes the signal insensitive to the film’s thermal conductivity. The dash-dotted curve repeats
the collapse boundary, and the white curve repeats
Eq.~\eqref{eq:bd_contact} from panel (a), showing that the region of good
conditioning ($A \lesssim 5$) extends only modestly beyond this boundary into the
contact-limited regime. Together, panels (a) and (b) show that 
the accessible parameter space is bounded by the interface- and
spreading-limited regimes. These are physical boundaries and not the instrument limitations that set the sensitivity limit of a point-contact measurement.

\subsection{\label{sec:app}Application: Thermal conductivity mapping}

Thermal conductivity mapping with nanoscale spatial resolution is essential for understanding and optimizing heat transport in nanoscale devices~\cite{pop2006heat,huxtable2004thermal}. In case of dielectric and interconnect thin films, processing and deposition conditions can introduce defects and structural inhomogeneities that locally modify thermal transport and lead to the formation of thermal hotspots, potentially degrading device performance~\cite{woon2025thermal,padovani2024dielectric}. SThM offers a powerful approach for probing such local variations in thermal transport with sub-$100$-nm spatial resolution.

Further simplification of Eq.~(\ref{eq:omega}) is possible for metallic films deposited on dielectric substrates, such as SiO$_2$ ($\kappa_{\mathrm{sub}}=1.38$ W\,m$^{-1}$K$^{-1}$). In the high-contrast limit $\Pi_3 \ll 1$, the spreading term collapses onto
the one-parameter master curve $\mathcal{G}(y)$ of Eq.~\eqref{eq:Gdef} (see Fig.~S2 and Table~S3 of the Supplementary Information). For
convenience in the inversion of Eq.~\eqref{eq:omega}, $\mathcal{G}$ is well
represented by the interpolation formula as given below,
\begin{equation}
\mathcal{G}(y) \simeq \left(1 + a\,y^{p}\right)^{-1},
\qquad a = 0.753,\quad p = 0.810,
\label{eq:Gfit}
\end{equation}
which reproduces Eq.~\eqref{eq:Gdef} to within $1.6\%$ error over
$0.3 \le y \le 300$. Equation~\eqref{eq:Gfit} is an interpolation, not a
scaling law: the exponent $p < 1$ mimics the logarithm of
Eq.~\eqref{eq:Gasym} over the fitting window and has no asymptotic meaning.
This interpolation formula departs from $\mathcal{G}$ increasingly outside that window. By considering Eq.~\eqref{eq:Gfit},
Eq.~\eqref{eq:omega} can be inverted explicitly to provide,

\begin{equation}
\mathcal{G}_i=\frac{4}{\Pi_{2,i}}
\left[\Omega_i-\frac{\Pi_{2,i}^{2}}{\pi}
\left(\frac{1}{\Lambda_c}+\frac{f_i}{\Lambda_{\mathrm{BD}}}\right)\right],
\label{eq:Gi}
\end{equation}
\begin{equation}
\boxed{\;\kappa_{\mathrm{film},i}
=\frac{\kappa_{\mathrm{sub}}\,b}{t}\left[\frac{1}{a}\left(\frac{1}{\mathcal{G}_i}-1\right)\right]^{1/p}\;}
\label{eq:closed}
\end{equation}
Here, $i$ denotes the individual pixel in the SThM signal map. In general, $f_i$ converges in two passes from an initial guess because it is a weak contribution to the total measured thermal resistance. Equation~(\ref{eq:closed}) provides a closed-form expression for pixel-by-pixel thermal conductivity mapping with SThM. With the known film thickness, together with a calibrated $h_c$ and the local contact radius, the local thermal conductivity can be determined directly.

We validate the framework against our previously reported SThM measurement on 14$\pm$1 nm thick Al film deposited on a SiO$_2$(1500~nm)/Si substrate~\cite{munde2026}. The closed-form inversion [Eq.~\eqref{eq:closed}] is applied to the experimentally measured SThM thermal-resistance map shown in Fig.~\ref{fig:kmap}(a). The tip-sample interfacial resistance, $R_{\mathrm{int}}=1.84$~MK~W$^{-1}$, is determined from the bare-oxide reference region. For simplicity, we take the contact radius $b$ to be equal to the tip apex
radius measured by scanning electron microscopy. Equation~(\ref{eq:closed}) then converts the measured $R_{ss}$ map into the corresponding spatially resolved thermal-conductivity map, $\kappa_{\mathrm{film}}$, shown in Fig.~\ref{fig:kmap}(b). Since the expression involves only elementary arithmetic operations, the inversion can be evaluated in a single vectorized pass and implemented directly in standard scanning-probe analysis software, such as Gwyddion. The resulting thermal-conductivity distribution (Fig.~\ref{fig:kmap} c) is approximately log-normal, with a median of 37.9~W~m$^{-1}$~K$^{-1}$. The distribution values are consistent with the previously reported mean thermal conductivity value of $\kappa_{\mathrm{Al}} = 45.1^{+4.7}_{-3.6}$~W~m$^{-1}$~K$^{-1}$ for the same film~\cite{munde2026}. As a stringent validation, Fig.~\ref{fig:kmap}(d) compares the closed-form solution with the full iterative inversion used in Ref.~\cite{munde2026}. The two solutions track each other pixel by pixel across the entire profile. 


Furthermore, the variation in the explicit inversion and full model is tested with the calibrated values of $\Lambda_c$ and $\Lambda_\mathrm{{BD}}$. Over the parameter range ($\Pi_2 = 0.28$,
$\kappa_{\mathrm{film}} = 14$-$280$\,W\,m$^{-1}$K$^{-1}$, i.e.
$y = 2.8$-$56$), the explicit inversion reproduces the full model to within
$3.4\%$, with $f$ iterated to convergence. The deviation is systematic and
negative, reflecting the combined error of the collapse and the 
interpolation formula [Eq.~\eqref{eq:Gfit}]. The error grows
outside this range, reaching $-8\%$ at $\kappa_{\mathrm{film}} = 7.7$ and
$-18\%$ at $4.6$\,W\,m$^{-1}$K$^{-1}$, where $y \lesssim 2$ and the collapse
no longer hold. The closed form is therefore
accurate and computationally efficient for routine quantitative SThM mapping
within the collapse regime; outside it,
$\kappa_{\mathrm{film}}$ should be obtained from the full model.

\section{\label{sec:conclusion}Conclusion}

We have shown that point-contact thermal-conductivity measurement of a thin
film is governed by a single invariant, $\Omega = R_{ss}\kappa_{\mathrm{sub}}t$,
and two control parameters, $\Pi_2 = t/b$ and
$\Pi_3 = \kappa_{\mathrm{sub}}/\kappa_{\mathrm{film}}$. The $(\Pi_2,\Pi_3)$ plane divides into a spreading-limited regime, where the
measured resistance carries information about $\kappa_{\mathrm{film}}$, and a
contact-limited regime, where it does not. Reducing the probe radius $b$ increases $\Pi_2$, and the contact term grows as
$\Pi_2^{2}$. Therefore, a sharper probe moves the measurement toward the
contact-limited regime. At high thermal contrast, the spreading resistance collapses onto a universal
function $\mathcal{G}(y)$ of the single variable
$y$. For $y \gg 1$, $\mathcal{G}(y)$
reduces to the logarithmic law of a conducting sheet, in which the film enters
only through its sheet conductance $\kappa_{\mathrm{film}}t$.

The central result is a measurability bound. We introduce the amplification
factor $A$, the factor by which a fractional error in the measured resistance
is magnified in the recovered thermal conductivity. Equation~(\ref{eq:A_asym}) shows
that $A$ grows in proportion to $R_{\mathrm{par}}\kappa_{\mathrm{film}}t$, so
sensitivity is lost once the parasitic resistance exceeds the film's sheet
thermal resistance. The amplification factor depends only on the contact geometry and interfaces. The instrument sets only the tolerable $A$, through its uncertainty in $R_{ss}$. As an example, consider the calibrated contact of Sec.~\ref{sec:app}
($R_{\mathrm{int}} = 1.84$\,MK\,W$^{-1}$, $b = 50$\,nm) on a $14\pm1$\,nm Al
film on SiO$_2$. For this geometry, the bound gives
$\kappa_{\mathrm{film}}^{\max} = 185$\,W\,m$^{-1}$K$^{-1}$ at $A_{\max} = 10$,
that is, the largest conductivity still recoverable when a $1\%$ error in
$R_{ss}$ becomes a $10\%$ error in $\kappa_{\mathrm{film}}$. Increasing the contact conductance is the only route that improves sensitivity without sacrificing spatial resolution. The 14$\pm$1 nm Al film on SiO$_2$ lies at $(\Pi_2,\Pi_3) = (0.28,\,0.031)$,
where the film contributes $26\%$ of the signal and $A = 5.2$. Because the regime boundaries depend only on $\Lambda_c$ and
$\Lambda_{\mathrm{BD}}$, the map can be redrawn for any film, substrate, and
probe, locating a measurement's limits. More broadly, the framework provides a quantitative basis for extracting film
properties from spatially resolved SThM measurements. It also allows one to
determine, before a measurement is made, whether a given probe and sample can
resolve them at all.
\\
\begin{acknowledgments}
We sincerely acknowledge Dr. Muhammad Ashraful Alam, Jai N. Gupta Distinguished Professor of Electrical and Computer Engineering at Purdue University, for his valuable guidance and insightful discussions that helped shape and advance this research. 
This work was supported in part by the Semiconductor Research Corporation (SRC) through the SPARC projects-based program in the CAST (Consortium for Advanced Sequential CFETs Pathfinding) track under the ``Nanoscale Thermal Metrology for Sequential CFETs'' project. 
The authors used Claude (Anthropic) to restructure and edit text drafted by the authors (Claude Opus 5), to critically review the derivations, and to develop the formal (Lean~4) and numerical verification described in Sec.~S6 of the Supplementary Information (Claude Opus 5.5 and Claude Fable~5.1). The authors reviewed all AI-assisted output and take full responsibility for the content of this paper.

\end{acknowledgments}

\section*{Data Availability Statement}
The experimental and modeling data that support the findings of this study are available from the corresponding author upon reasonable request.

\section*{References}

\bibliography{aipsamp}

\clearpage
\onecolumngrid
\setcounter{equation}{0}
\setcounter{figure}{0}
\setcounter{table}{0}
\renewcommand{\theequation}{S.\arabic{equation}}
\renewcommand{\thefigure}{S\arabic{figure}}
\renewcommand{\thetable}{S\arabic{table}}
\begin{center}
{\large\bfseries Supplementary Information}\\[4pt]
{\large\bfseries Universal scaling and measurability limits of thermal-conductivity measurement by a nanoscale point contact}\\[6pt]
Ram Munde and Raisul Islam
\end{center}
\section*{S1. Buckingham $\Pi$ Theorem}
Consider n physically relevant dimensional variables, $Q_1,Q_2,\ldots,Q_n$, that are related by an unknown set of dimensionally homogeneous equations (Eq. \ref{eq:S_F}). We follow the formalism and the notation of Buckingham's original paper.\cite{buckingham1914physically}

\begin{equation}
F(Q_1,Q_2,\ldots,Q_n)=0,
\label{eq:S_F}
\end{equation}
is reducible by dimensional homogeneity to
\begin{equation}
\psi(\Pi_1,\Pi_2,\ldots,\Pi_i)=0,
\label{eq:S_psi}
\end{equation}
where each $\Pi$ is a dimensionless product
\begin{equation}
\Pi=Q_1^{a_1}Q_2^{a_2}\cdots Q_n^{a_n},
\label{eq:S_prod}
\end{equation}
and the number of independent products is
\begin{equation}
i=n-k,
\label{eq:S_ink}
\end{equation}
with $k$ the number of arbitrary fundamental units needed as a basis for the absolute system by
which the $Q$'s are measured. Two points of Buckingham's treatment are used
repeatedly below and are worth stating at the outset.

\begin{enumerate}
\item[(i)] Buckingham is
explicit that the $k$ fundamental units ``need not be $[m,l,t]$'': in his screw-propeller example
$k=3$, but the triad may be taken as $[F,\rho,D]$ or $[\rho,D,S]$ instead of
$[m,l,t]$, while other combinations ``could not be used.'' The value of $k$ must therefore be
determined from the quantities of the problem, not assumed from the number of physical dimensions
that appear. In our problem, this distinction is decisive: four dimensional symbols appear, yet
$k=2$ (Sec.~S2).
\item[(ii)] \emph{Any function of a $\Pi$ is itself a valid $\Pi$.} Since the $\Pi$'s enter
Eq.~(\ref{eq:S_psi}) only as arguments of an indeterminate function and are subject only to being
dimensionless, once a particular product has been found, one is ``at liberty to replace this by any
function of it.'' We use this freedom in Sec.~S3 to pass between equivalent sets of groups, and in the main text to form the measured invariant $\Omega$ as a product of three
$\Pi$'s.
\end{enumerate}

\section*{\label{sec:s2}S2. The physical quantities and their dimensions}

The measured solid-solid thermal resistance of the tip-sample contact is modeled as a series
network as given below \cite{munde2026} (Eq.\ref{eq:S_network}),
\begin{equation}
R_{ss}=R_{\mathrm{int}}(h_c,b)+R_{\mathrm{spr}}(t,b,\kappa_{\mathrm{film}},\kappa_{\mathrm{sub}})+f(t,b,\kappa_{\mathrm{film}},\kappa_{\mathrm{sub}})\,R_{\mathrm{BD}}(b,h_{\mathrm{BD}})
\label{eq:S_network}
\end{equation}
with $R_{\mathrm{BD}}=1/(h_{\mathrm{BD}}\pi b^{2})$ and $R_{\mathrm{int}}=1/(h_c\pi b^{2})$. 

The relation therefore involves $n=7$ kinds of quantity, listed in Table~\ref{tab:S_dims}. 

\begin{table}[t]
\caption{The $n=7$ kinds of quantity entering the tip-sample contact relation, with their
dimensions in the $[m,l,t,\theta]$ system. $R_{ss}$ is the dependent quantity.}
\label{tab:S_dims}
\begin{ruledtabular}
\begin{tabular}{llll}
Name & Symbol & Dimensions & Exponents $(m,l,t,\theta)$ \\
\colrule
SThM tip-sample solid-solid resistance & $R_{ss}$ (K\,W$^{-1}$) & $[m^{-1}l^{-2}t^{3}\theta]$ & $(-1,-2,\;3,\;1)$ \\
Film thermal conductivity & $\kappa_{\mathrm{film}}$(W\,m$^{-1}$K$^{-1}$) & $[m\,l\,t^{-3}\theta^{-1}]$ & $(\;1,\;1,-3,-1)$ \\
Substrate thermal conductivity & $\kappa_{\mathrm{sub}}$(W\,m$^{-1}$K$^{-1}$) & $[m\,l\,t^{-3}\theta^{-1}]$ & $(\;1,\;1,-3,-1)$ \\
Contact radius & $b$ (m) & $[l]$ & $(\;0,\;1,\;0,\;0)$ \\
Film thickness & $t$ (m) & $[l]$ & $(\;0,\;1,\;0,\;0)$ \\
Buried-interface conductance & $h_{\mathrm{BD}}$ (W\,m$^{-2}$K$^{-1}$) & $[m\,t^{-3}\theta^{-1}]$ & $(\;1,\;0,-3,-1)$ \\
Tip-sample interfacial conductance & $h_c$ (W\,m$^{-2}$K$^{-1}$) & $[m\,t^{-3}\theta^{-1}]$ & $(\;1,\;0,-3,-1)$ \\
\end{tabular}
\end{ruledtabular}
\end{table}

The dimensional matrix formed from the last column of Table~\ref{tab:S_dims} is
\begin{equation}
\mathsf{D}=
\bordermatrix{
 & R_{ss} & \kappa_{\mathrm{film}} & \kappa_{\mathrm{sub}} & b & t & h_{\mathrm{BD}} & h_c \cr
m & -1 & 1 & 1 & 0 & 0 & 1 & 1 \cr
l & -2 & 1 & 1 & 1 & 1 & 0 & 0 \cr
t & \;\;3 & -3 & -3 & 0 & 0 & -3 & -3 \cr
\theta & \;\;1 & -1 & -1 & 0 & 0 & -1 & -1 \cr}.
\label{eq:S_matrix}
\end{equation}
The rank of this matrix  $\mathsf{D}$ is $2$. Two independent
verifications follow.

\paragraph*{(a) Row dependencies.} The above matrix shows that,
\begin{equation}
\mathsf{D}_{t}=-3\,\mathsf{D}_{m},\qquad \mathsf{D}_{\theta}=-\,\mathsf{D}_{m},
\end{equation}
so the $t$ and $\theta$ rows are proportional to the $m$ row, and only the $m$ and $l$ rows are
independent; hence $\mathrm{rank}(\mathsf{D})=2$. Physically, every quantity in
Table~\ref{tab:S_dims} is a thermal resistance, a conductivity, a conductance, or a length, and
$m$, $t$, and $\theta$ occur in these only through the fixed combination $\theta/(ml^{2}t^{-3})
=\theta W^{-1}$. The problem is thus expressed on the two-symbol basis $\{L,\;\Theta W^{-1}\}$:
resistances scale as $\Theta W^{-1}$, conductivities as $L^{-1}(\Theta W^{-1})^{-1}$, and
conductances as $L^{-2}(\Theta W^{-1})^{-1}$.

\paragraph*{(b) An explicit admissible pair.} Following Buckingham's remark (i), we exhibit two of
the $Q$'s that suffice as fundamental units. The columns for $\kappa_{\mathrm{film}}$ and $b$,
\begin{equation}
\bm{q}_{\kappa}=(1,1,-3,-1)^{\!\top},\qquad \bm{q}_{b}=(0,1,0,0)^{\!\top},
\end{equation}
are linearly independent, and every remaining column is a linear combination of them:
\begin{equation}
\bm{q}_{R}=-\bm{q}_{\kappa}-\bm{q}_{b},\quad
\bm{q}_{\kappa_{\rm sub}}=\bm{q}_{\kappa},\quad
\bm{q}_{t}=\bm{q}_{b},\quad
\bm{q}_{h}=\bm{q}_{\kappa}-\bm{q}_{b},
\end{equation}
the last holding for both $h_{\mathrm{BD}}$ and $h_c$. Hence, $[\kappa_{\mathrm{film}}]$ and $[b]$ form a
complete basis for measuring all seven quantities, and
\begin{equation}
\boxed{\;k=2,\qquad i=n-k=7-2=5.\;}
\label{eq:S_count}
\end{equation}

\paragraph*{Remark on the counting.} Equation~(\ref{eq:S_count}) states that five independent
dimensionless products exist among the seven quantities. One of these necessarily contains the
dependent quantity $R_{ss}$; solving Eq.~(\ref{eq:S_psi}) for it leaves \emph{four} independent
arguments. Equivalently, and more directly: $R_{ss}$ is a function of the remaining six
quantities, and those six, on a rank-two basis, yield $6-2=4$ groups, to which the dependent group is
added. The two statements are identical, and both are used in the main text.

Now, we apply Buckingham's Eq.(\ref{eq:S_prod}),
\begin{equation}
[\Pi]=[Q_1^{\alpha}Q_2^{\beta}\cdots Q_k^{\lambda}P]=[1],
\label{eq:S_bucket}
\end{equation}
taking as the two repeating quantities
\begin{equation}
Q_1=\kappa_{\mathrm{film}},\qquad Q_2=b,
\end{equation}
which were shown in Sec.~S2(b) to be an admissible pair, and letting the remaining five quantities
play the role of the $P$'s in turn:
\begin{equation}
P_1=R_{ss},\quad P_2=t,\quad P_3=\kappa_{\mathrm{sub}},\quad P_4=h_{\mathrm{BD}},\quad P_5=h_c .
\end{equation}
Each product has the form $\Pi=\kappa_{\mathrm{film}}^{\alpha}b^{\beta}P$, and substituting the
dimensions of Table~\ref{tab:S_dims} gives
\begin{equation}
\big[m^{\alpha}l^{\alpha}t^{-3\alpha}\theta^{-\alpha}\big]\big[l^{\beta}\big]\big[P\big]=[1],
\label{eq:S_general}
\end{equation}
from which the exponent of each of $m$, $l$, $t$, and $\theta$ must vanish separately.

\begin{enumerate}
 
\item  $P_1=R_{ss}$: the dependent group

With $[R_{ss}]=[m^{-1}l^{-2}t^{3}\theta]$, Eq.~(\ref{eq:S_general}) reads
\begin{equation}
m^{\alpha-1}\,l^{\alpha+\beta-2}\,t^{-3\alpha+3}\,\theta^{-\alpha+1}=[1],
\end{equation}
giving the four conditions
\begin{equation}
\alpha-1=0,\qquad \alpha+\beta-2=0,\qquad -3\alpha+3=0,\qquad -\alpha+1=0 .
\end{equation}
The first, third, and fourth are the same equation, $\alpha=1$ (a direct consequence of
$\mathrm{rank}=2$: only two of the four conditions are independent). The second then gives
$\beta=1$, and
\begin{equation}
\boxed{\;\Pi_1=R_{ss}\,\kappa_{\mathrm{film}}\,b\;}
\label{eq:S_Pi1}
\end{equation}

\item $P_2=t$: the confinement ratio

With $[t]=[l]$,
\begin{equation}
m^{\alpha}\,l^{\alpha+\beta+1}\,t^{-3\alpha}\,\theta^{-\alpha}=[1]
\;\Longrightarrow\; \alpha=0,\;\;\beta=-1,
\end{equation}
so that
\begin{equation}
\boxed{\;\Pi_2=\frac{t}{b}\;}
\label{eq:S_Pi2}
\end{equation}

\item $P_3=\kappa_{\mathrm{sub}}$: the conductivity mismatch

With $[\kappa_{\mathrm{sub}}]=[m\,l\,t^{-3}\theta^{-1}]$,
\begin{equation}
m^{\alpha+1}\,l^{\alpha+\beta+1}\,t^{-3\alpha-3}\,\theta^{-\alpha-1}=[1]
\;\Longrightarrow\; \alpha=-1,\;\;\beta=0,
\end{equation}
so that
\begin{equation}
\boxed{\;\Pi_3=\frac{\kappa_{\mathrm{sub}}}{\kappa_{\mathrm{film}}}\;}
\label{eq:S_Pi3}
\end{equation}
The contact radius does not enter: the conductivity mismatch is a pure ratio, as it must be.

\item $P_4=h_{\mathrm{BD}}$: the buried-interface group

With $[h_{\mathrm{BD}}]=[m\,t^{-3}\theta^{-1}]$ (note the vanishing exponent of $l$),
\begin{equation}
m^{\alpha+1}\,l^{\alpha+\beta}\,t^{-3\alpha-3}\,\theta^{-\alpha-1}=[1]
\;\Longrightarrow\; \alpha=-1,\;\;\beta=+1,
\end{equation}
so that
\begin{equation}
\boxed{\;\Pi_4=\frac{h_{\mathrm{BD}}\,b}{\kappa_{\mathrm{film}}}\;}
\label{eq:S_Pi4}
\end{equation}

\item $P_5=h_c$: the tip-sample interface group

The dimensions of $h_c$ are those of $h_{\mathrm{BD}}$, so the same exponents follow, and
\begin{equation}
\boxed{\;\Pi_5=\frac{h_c\,b}{\kappa_{\mathrm{film}}}\;}
\label{eq:S_Pi5}
\end{equation}
The interface resistance itself is recovered as a derived group,
\begin{equation}
\Pi_{\mathrm{int}}\equiv R_{\mathrm{int}}\kappa_{\mathrm{film}}b
=\frac{\kappa_{\mathrm{film}}b}{h_c\pi b^{2}}\cdot\frac{1}{1}
=\frac{1}{\pi\Pi_5},
\label{eq:S_Piint}
\end{equation}
which is legitimate by Buckingham's remark (ii), being a function of $\Pi_5$ alone.
\end{enumerate}

The five products of $\Pi$ groups are collected in Table~\ref{tab:S_exponents} as rows of the exponent
matrix $\mathsf{A}$, whose columns are the seven quantities. Each row of $\mathsf{A}$ contains a
$+1$ in the column of its own $P$ and zeros in the columns of the other four $P$'s; the
$5\times5$ submatrix formed by the $P$ columns is therefore the identity, $\mathsf{A}$ has rank
$5$, and the five products are independent. Since $i=n-k=5$ by Eq.~(\ref{eq:S_count}), the set is
also complete: no further independent dimensionless product can be formed.

\begin{table}[t]
\caption{Exponent matrix of the five independent dimensionless products. Each row gives the
exponents $a_j$ in $\Pi=\prod_j Q_j^{a_j}$.}
\label{tab:S_exponents}
\begin{ruledtabular}
\begin{tabular}{lccccccc}
& $R_{ss}$ & $\kappa_{\mathrm{film}}$ & $\kappa_{\mathrm{sub}}$ & $b$ & $t$ & $h_{\mathrm{BD}}$ & $h_c$ \\
\colrule
$\Pi_1=R_{ss}\kappa_{\mathrm{film}}b$        & 1 & \;\;1 & 0 & \;\;1 & 0 & 0 & 0 \\
$\Pi_2=t/b$                             & 0 & \;\;0 & 0 & $-1$  & 1 & 0 & 0 \\
$\Pi_3=\kappa_{\mathrm{sub}}/\kappa_{\mathrm{film}}$ & 0 & $-1$ & 1 & \;\;0 & 0 & 0 & 0 \\
$\Pi_4=h_{\mathrm{BD}}b/\kappa_{\mathrm{film}}$ & 0 & $-1$ & 0 & \;\;1 & 0 & 1 & 0 \\
$\Pi_5=h_cb/\kappa_{\mathrm{film}}$          & 0 & $-1$ & 0 & \;\;1 & 0 & 0 & 1 \\
\end{tabular}
\end{ruledtabular}
\end{table}

By Eq.~(\ref{eq:S_psi}) the complete relation is therefore
\begin{equation}
\psi(\Pi_1,\Pi_2,\Pi_3,\Pi_4,\Pi_5)=0,
\end{equation}
and solving for the single group that contains the dependent quantity,
\begin{equation}
\Pi_1=\mathcal{F}(\Pi_2,\Pi_3,\Pi_4,\Pi_5),
\label{eq:S_master}
\end{equation}
an unknown function of four arguments.

\section*{S3. Equivalent bases and the constants $\Lambda$}

Buckingham notes that a different admissible set of fundamental quantities gives an equally valid
set of products, which, while essentially equivalent, presents a different
appearance. Two alternative choices are used in the main text.

\paragraph*{Substrate normalization.} Taking $Q_1=\kappa_{\mathrm{sub}}$, $Q_2=b$, which is also an admissible
pair since $\bm{q}_{\kappa_{\rm sub}}=\bm{q}_{\kappa}$, the same procedure gives
\begin{equation}
\Pi_1'=R_{ss}\kappa_{\mathrm{sub}}b,\quad
\Pi_3'=\frac{\kappa_{\mathrm{film}}}{\kappa_{\mathrm{sub}}}=\Pi_3^{-1},\quad
\Pi_4'=\frac{h_{\mathrm{BD}}b}{\kappa_{\mathrm{sub}}},\quad
\Pi_5'=\frac{h_cb}{\kappa_{\mathrm{sub}}} .
\end{equation}
These are related to the previous set by $\Pi_1'=\Pi_1\Pi_3$, $\Pi_4'=\Pi_4/\Pi_3$, and
$\Pi_5'=\Pi_5/\Pi_3$, i.e., by products of $\Pi$'s, and are therefore admissible by remark (ii).
Their practical advantage is that $\Pi_4'$ and $\Pi_5'$ contain only known quantities.

\paragraph*{Thickness normalization.} Taking $Q_1=\kappa_{\mathrm{sub}}$, $Q_2=t$ provides the two
constants used throughout the main text.
\begin{equation}
\boxed{\;\Lambda_{\mathrm{BD}}=\frac{h_{\mathrm{BD}}\,t}{\kappa_{\mathrm{sub}}},\qquad
\Lambda_{c}=\frac{h_c\,t}{\kappa_{\mathrm{sub}}}\;}
\label{eq:S_lambdas}
\end{equation}
For a fixed film thickness and a calibrated probe, these numbers are fixed once for the entire
experiment, and they are related to the groups of Sec.~S2 by
\begin{equation}
\Lambda_{\mathrm{BD}}=\Pi_4\frac{\Pi_2}{\Pi_3},\qquad
\Lambda_{c}=\Pi_5\frac{\Pi_2}{\Pi_3},
\label{eq:S_lam_rel}
\end{equation}
Again, products of $\Pi$.

\section*{S4. Dimensionless form of the contact model}

The spreading resistance solution  given by  Dryden~\cite{dryden1983} and the Yovanovich-Muzychka~\cite{yovanovich2002analytical, muzychka1999thermal} for a film on half-space 
is,
\begin{equation}
R_{\mathrm{spr}} = \frac{1}{\pi \kappa_{\mathrm{film}} b} 
\int_{0}^{\infty} 
\left[
\frac{1 + K \exp(-2\xi t / b)}
{1 - K \exp(-2\xi t / b)}
\right]
J_1(\xi)\,\sin(\xi)\,
\frac{d\xi}{\xi^2},
\label{eq:rspr_SI}
\end{equation}

where $K$ is the thermal-conductivity mismatch parameter, defined as
\begin{equation}
K = \frac{1-\kappa_{\mathrm{sub}}/\kappa_{\mathrm{film}}}
{1+\kappa_{\mathrm{sub}}/\kappa_{\mathrm{film}}},
\end{equation}
The integration variable $\xi$ is dimensionless; the exponential argument is $2\xi\Pi_2$ and
$K=(1-\Pi_3)/(1+\Pi_3)$. Hence, the entire integral depends only on $\Pi_2$ and $\Pi_3$, and
$\Psi(\Pi_2, \Pi_3)$ is therefore a dimensionless group of the same form as $\Pi_1$ with $R_{\mathrm{spr}}$ in place of $R_{ss}$.
\begin{equation}
\Psi(\Pi_2,\Pi_3)\equiv R_{\mathrm{spr}}\kappa_{\mathrm{film}}b
=\frac{1}{\pi}\int_0^{\infty}\frac{1+K e^{-2\xi\Pi_2}}{1-K e^{-2\xi\Pi_2}}
\frac{J_1(\xi)\sin\xi}{\xi^{2}}\,d\xi
\label{eq:S_psi_def}
\end{equation}
The geometric dilution factor, with the lateral spreading length
$L_{\mathrm{spread}}=t\sqrt{\kappa_{\mathrm{film}}/\kappa_{\mathrm{sub}}}$, is already dimensionless and is
expressed in the same two groups:
\begin{equation}
f=\left(\frac{b}{b+L_{\mathrm{spread}}}\right)^{2}
=\left(1+\frac{t}{b}\sqrt{\frac{\kappa_{\mathrm{film}}}{\kappa_{\mathrm{sub}}}}\right)^{-2}
=\left(1+\Pi_2\Pi_3^{-1/2}\right)^{-2}.
\label{eq:S_f}
\end{equation}
Dividing Eq.~(\ref{eq:S_network}) by $1/(\kappa_{\mathrm{film}}b)$ and using
Eqs.~(\ref{eq:S_Piint}), (\ref{eq:S_psi_def}), and (\ref{eq:S_f}), along with
$R_{\mathrm{BD}}\kappa_{\mathrm{film}}b=\kappa_{\mathrm{film}}/(h_{\mathrm{BD}}\pi b)=1/(\pi\Pi_4)$, gives the
explicit form of Eq.~(\ref{eq:S_master}).
\begin{equation}
\boxed{\;\Pi_1=\frac{1}{\pi\Pi_5}+\Psi(\Pi_2,\Pi_3)+\frac{f(\Pi_2,\Pi_3)}{\pi\Pi_4}\;}
\label{eq:S_master_explicit}
\end{equation}

\begin{figure}[htbp]
\includegraphics[width=0.95\textwidth]{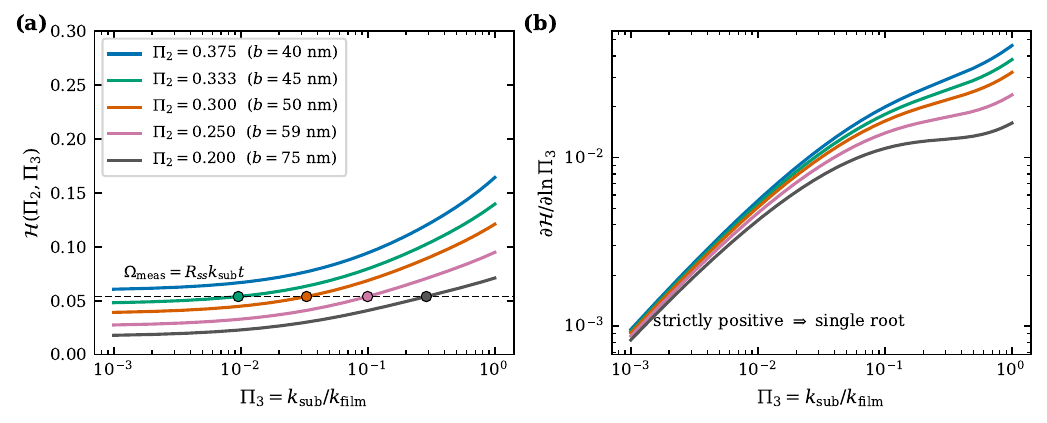}
\caption{Uniqueness of the per-pixel inversion. (a) The master surface
$\mathcal{H}(\Pi_2,\Pi_3)$ as a function of $\Pi_3$ at five fixed
values of $\Pi_2$. (b) The logarithmic derivative
$\partial\mathcal{H}/\partial\ln\Pi_3$ is strictly positive over three decades of $\Pi_3$
(minimum $8\times10^{-4}$), establishing the monotonicity that guarantees the single crossing
in (a).}
\label{fig:S_mono}
\end{figure}
\begin{figure}[htbp]
\includegraphics[width=0.95\textwidth]{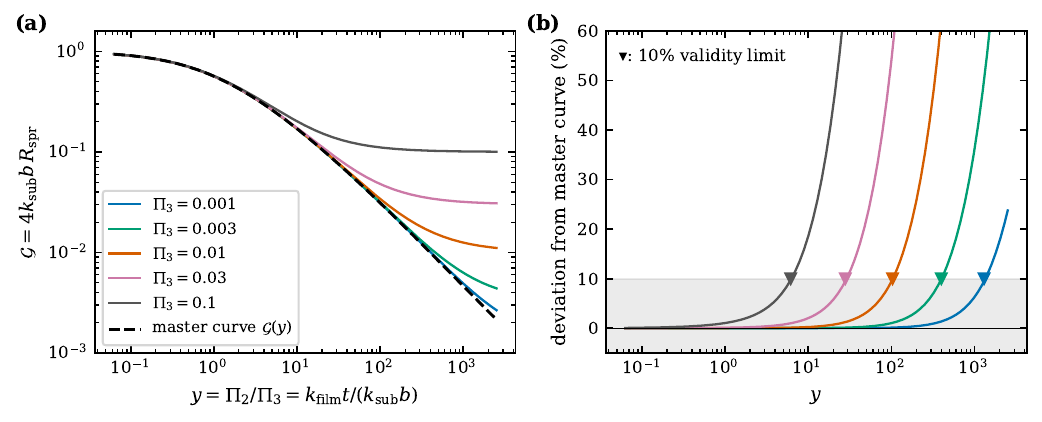}
\caption{Collapse onto a master curve for different values of $\Pi_3$. (a) The curves collapse onto a single master curve in the limit $\Pi_3\ll1$. (b) The 10\% error window defines a practical range of $y$ applicable to a broad range of thin metallic films.
}
\label{fig:S_collapse}
\end{figure}
\newpage
\section*{S5. High-contrast expansion of the spreading kernel}
\label{sec:S_collapse}

The spreading function [Eq.~(10) of the main text] contains the kernel
\begin{equation}
\mathcal{K}(\xi) = \frac{1+Ke^{-2\xi\Pi_2}}{1-Ke^{-2\xi\Pi_2}},
\qquad
K = \frac{1-\Pi_3}{1+\Pi_3} = 1 - \frac{2\Pi_3}{1+\Pi_3}.
\label{eq:S_kernel}
\end{equation}
For $\Pi_3 \ll 1$, $K \simeq 1 - 2\Pi_3$, with a relative error in $1-K$ of the order
$\Pi_3$. For $\xi\Pi_2 \ll 1$, $e^{-2\xi\Pi_2} \simeq 1 - 2\xi\Pi_2$. Because
the weight $J_1(\xi)\sin\xi/\xi^2$ is appreciable only for $\xi \lesssim 1$,
this second condition requires $\Pi_2 \lesssim 1$. Since both $K$ and
$e^{-2\xi\Pi_2}$ are close to unity, the denominator of
Eq.~\eqref{eq:S_kernel} is controlled by their deviations from unity.
Retaining these to first order,
\begin{equation}
1 - Ke^{-2\xi\Pi_2}
\simeq 2\Pi_3 + 2\xi\Pi_2
= 2\Pi_3\left(1 + y\xi\right),
\qquad y = \frac{\Pi_2}{\Pi_3},
\label{eq:S_denom}
\end{equation}
where the second-order product $4\Pi_3\xi\Pi_2$ has been neglected. Neither
first-order term may be dropped relative to the other, since their ratio
$y\xi$ is unrestricted. With the numerator $\simeq 2$,
$\mathcal{K} \simeq [\Pi_3(1+y\xi)]^{-1}$, and
\begin{equation}
4\Pi_3\Psi \simeq
\frac{4}{\pi}\int_0^{\infty}
\frac{J_1(\xi)\sin\xi}{\xi^{2}(1+y\xi)}\,d\xi
\equiv \mathcal{G}(y),
\label{eq:S_G}
\end{equation}
with $\mathcal{G}(0) = 1$, recovering the substrate half-space limit
$\Psi = 1/(4\Pi_3)$.

Physically, $\xi$ is a radial wavenumber in units of $1/b$. The term $1$ in
$1 + y\xi$ represents conduction into the substrate, with surface conductance
$\propto \kappa_{\mathrm{sub}}\xi/b$, and $y\xi$ represents lateral conduction
through the film, with conductance $\propto \kappa_{\mathrm{film}}t(\xi/b)^2$.
In the collapse regime, the film acts as a conducting sheet, and
$\kappa_{\mathrm{film}}$ and $t$ enter only through the sheet conductance
$\kappa_{\mathrm{film}}t$.

Both conditions are necessary. Table~\ref{tab:S_collapse} compares
Eq.~\eqref{eq:S_G} with the full $\Psi$. At fixed $\Pi_3 = 0.033$, the error
grows from $2\%$ to $62\%$ as $y$ increases because $\Pi_2 = y\Pi_3$ exceeds
unity and the exponential expansion fails.

\begin{table}[h]
\caption{Relative excess of the full $4\Pi_3\Psi$ over $\mathcal{G}(y)$, i.e.\ $4\Pi_3\Psi/\mathcal{G}-1$.}
\label{tab:S_collapse}
\begin{ruledtabular}
\begin{tabular}{cccc}
 & $y = 9.1$ & $y = 30$ & $y = 100$ \\
\hline
$\Pi_3 = 0.033$ & $+2.1\%$ ($\Pi_2 = 0.30$) & $+12.4\%$ ($\Pi_2 = 0.99$) & $+62\%$ ($\Pi_2 = 3.3$) \\
$\Pi_3 = 0.10$  & $+16\%$ ($\Pi_2 = 0.91$) & $+72\%$ ($\Pi_2 = 3.0$) & $+256\%$ ($\Pi_2 = 10$) \\
\end{tabular}
\end{ruledtabular}
\end{table}

\section*{S6. Verification of the derivations}

The dimensional analysis and the algebraic results of this work were formally verified in the Lean~4 proof assistant (v4.35.0-rc2 with the Mathlib library). The formalization contains 46 theorems. It covers the row dependencies and the rank of the dimensional matrix and the independence of $\Pi_1$-$\Pi_5$ (Sec.~S2), Eqs.~(8), (9), (11), (12), (18), (23) and (24) of the main text, and the expansion of the kernel denominator used for the collapse (Sec.~S5). It also covers the amplification factor in the collapse regime and its asymptotic form [Eqs.~(27) and (28) of the main text], the resolvable conductivity bound [Eq.~(29)], the explicit inversion [Eqs.~(32) and (33)], and the dependence of $A$ on $h_c$, $b$, and $t$ that underlies the design rule [Eq.~(30)]. The solvability condition [Eq.~(25)] is proved on the condition that $\mathcal{H}$ increases monotonically in $\Pi_3$ and that $\Pi_3\Psi \to 0$ approaches $\Pi_3 \to 0$, both of which are verified numerically as described below. The project builds without unproven statements (\texttt{sorry}) and uses no axioms beyond the three standard axioms of Lean.

Results that depend on the Bessel integral of Eq.~(10) of the main text cannot be formalized because Mathlib does not include Bessel functions. These include Eqs.~(14)-(17), the constant $C$, the monotonicity of $\mathcal{H}$ in $\Pi_3$, and all values obtained by quadrature. They were verified numerically with scipy and mpmath quadrature, and independently with rigorous interval arithmetic (Arb, through python-flint). The monotonicity of $\mathcal{H}$ in $\Pi_3$ was certified in this way over $10^{-3} \leq \Pi_2 \leq 10$ and $10^{-6} \leq \Pi_3 \leq 10^{2}$.

Claude Opus 5 (Anthropic) was also used to assist in deriving the interpolation formula for $\mathcal{G}(y)$ [Eq.~(31) of the main text] and to write the Python scripts that evaluate the expressions of Sec.~III.B of the main text. All AI-assisted derivations and numerical results were checked by the authors against the governing equations.

The Lean formalization and the numerical verification scripts were developed with the assistance of Claude Opus 5.5 (Anthropic). The independent numerical verification used Claude Fable~5.1. All statements and results were reviewed by the authors. The Lean source and the verification scripts are available from the corresponding author upon reasonable request.

\end{document}